\documentclass{emulateapj}
\usepackage{apjfonts}
\usepackage{xspace}
\usepackage{amsmath}
\usepackage{hyperref}
\usepackage{epsf}
\usepackage{graphicx}
\usepackage{epstopdf}

\def\d{{\rm d}}
\def\p{{\rm p}}
\def\H{{\rm H}}

\def\sl{{\rm sl}}
\def\g{{\rm gas}}

\begin{document}
\shorttitle{Gas Clumping in Cluster Outskirts}
\shortauthors{Nagai \& Lau}

\title{Gas Clumping in the Outskirts of $\Lambda$CDM Clusters}

\author{
Daisuke Nagai\altaffilmark{1},
Erwin T. Lau\altaffilmark{2}
}

\altaffiltext{1}{Department of Physics, Yale University, New Haven, CT 06520, U.S.A.; {daisuke.nagai@yale.edu}} 
\altaffiltext{2}{Key Laboratory for Research in Galaxies and Cosmology, Shanghai Astronomical Observatory; The Partner Group of MPA; 80 Nandan Road, Shanghai 200030, China}  
     
% User-supplied List of keywords.
\keywords{cosmology: theory --- galaxies: clusters: general --- X-rays: galaxies: clusters --- methods: numerical}

\begin{abstract}
Recent {\it Suzaku} X-ray observations revealed that the observed entropy profile of the intracluster medium (ICM) deviates significantly from the prediction of hydrodynamical simulations of galaxy clusters. In this work, we show that gas clumping introduces significant biases in X-ray measurements of the ICM profiles in the outskirts of galaxy clusters. Using hydrodynamical simulations of galaxy cluster formation in a concordance $\Lambda$CDM model, we demonstrate that gas clumping leads to an overestimate of the observed gas density and causes flattening of the entropy profile.  Our results suggest that gas clumping must be taken into account when interpreting X-ray measurements of cluster outskirts. 
\\
\end{abstract}

%-------------------------
\section{Introduction}
\label{sec:intro}
%-------------------------

In recent years, galaxy clusters have emerged as one of the unique and powerful laboratories for cosmology and astrophysics. Being the largest and most magnificent structures in the Universe, clusters of galaxies serve as excellent tracers of the growth of cosmic structures. 
Current generation of X-ray cluster surveys have provided independent confirmation of cosmic acceleration and significantly tighten constraints on the nature of dark energy \citep{allen_etal08,vikhlinin_etal09} and alternative theories of gravity \citep[e.g.,][]{schmidt_etal09}. New X-ray survey missions (e.g., {\it eROSITA}) are underway to improve current cosmological constraints. 

Outskirts of galaxy clusters have special importance in cluster cosmology, because they are believed to be much less susceptible to complicated cluster astrophysics, such as radiative gas cooling, star formation, and energy injection from active galactic nuclei. Dominant physical processes in the outskirts are limited to the gravity-driven collisionless dynamics of dark matter and hydrodynamics of the intracluster medium (ICM). However, until very recently, observational studies of the ICM have been limited to radii considerably smaller than the virial radius of clusters. 

Recently, X-ray observations with {\it Suzaku} have extended measurements of the ICM profile out to and beyond the virial radius for a number of clusters \citep{bautz_etal09,george_etal09,reiprich_etal09,hoshino_etal10,kawaharada_etal10,simionescu_etal11}. While measurement uncertainties are still large, the initial results indicate that the observed ICM profiles deviate significantly from the predictions of hydrodynamical cluster simulations \citep[e.g.,][]{george_etal09}. In addition to testing model of structure formation, these new measurements are important for controlling systematic uncertainties for cluster-based cosmological measurements.

In the hierarchical structure formation model, clusters grow by accreting materials from the surrounding large-scale structure in their outer envelope.  Numerical simulations predict that accretion and mergers are ubiquitous and important for cluster formation, and the accretion physics gives rise to internal gas motions and inhomogeneous gas density distribution ("clumpiness") in the ICM. Recent hydrodynamical simulations, for example, indicate that non-thermal pressure provided by gas motions significantly modifies the ICM profiles in the cluster outskirts \citep[e.g.,][]{iapichino_niemeyer08,vazza_etal09,lau_etal09}. Gas clumping, on the other hand, also introduces biases the X-ray measurements of the ICM. \citet{mathiesen_etal99}, for example, pointed out that gas clumping can lead to an overestimate of the ICM mass by $\sim 10\%$ around half of the virial radius . However, the impact of gas clumping on cluster outskirts has received little attention in the literature.

In this work, we investigate effects of gas clumping on X-ray measurements in the outskirts of galaxy clusters. Using hydrodynamical cluster simulations, we show that gas clumping leads to an overestimate of gas density and causes flattening of the entropy profile in cluster outskirts ($r \gtrsim r_{200}$). Our results suggest that gas clumping should be treated properly when interpreting X-ray measurements in the envelop of galaxy clusters. 

The paper is organized as follows. We describe our simulations in Section~\ref{sec:sim} and define the clumping factor in Section~\ref{sec:clumping}.  In Section~\ref{sec:results} we present our results. Our main findings are summarized in Section~\ref{sec:summary}.

%------------------------
\section{Simulations}
\label{sec:sim}
%------------------------

In this work, we analyze a sample of 16 simulated groups and clusters of galaxies. The simulations presented here are presented in \citet{nagai_etal07a} and \citet{nagai_etal07b}, and we refer the reader to these papers for more details. We briefly summarize relevant parameters of the simulations here.

These simulations are performed using the Adaptive Refinement Tree (ART) $N$-body$+$gas-dynamics code \citep{kra99,kra02}, which is an Eulerian code that uses adaptive refinement in space and time, and non-adaptive refinement in mass \citep{klypin_etal01} to achieve the dynamic ranges to resolve the cores of halos formed in self-consistent cosmological simulations. The simulations assume a flat {$\Lambda$}CDM model: $\Omega_{\rm m}=1-\Omega_{\Lambda}=0.3$, $\Omega_{\rm b}=0.04286$, $h=0.7$ and $\sigma_8=0.9$, where the Hubble constant is defined as $100h{\ \rm km\ s^{-1}\ Mpc^{-1}}$, and $\sigma_8$ is the mass variance within spheres of radius $8\,h^{-1}$~Mpc.  The simulations were run using a uniform $128^3$ grid with 8 levels of mesh refinement.  The box size for CL101--CL107 is $120\,h^{-1}$~Mpc comoving on a side and is $80\,h^{-1}$~Mpc comoving for CL3--CL24. This corresponds to peak spatial resolution of $\approx 7\,h^{-1}$~kpc and $5\,h^{-1}$~kpc for the two box sizes respectively.  Only the inner regions $\sim 3-10\,h^{-1}$~Mpc surrounding the cluster center were adaptively refined. The dark matter (DM) particle mass in the region around each cluster was $m_{\rm p} \simeq 9.1\times 10^{8}\,h^{-1}\, {\rm M_{\odot}}$ for CL101--107 and $ m_{\rm p} \simeq 2.7\times 10^{8}\,h^{-1}\,{\rm M_{\odot}}$ for CL3--24, while other regions were simulated with a bottom mass resolution. 

To assess the effects of galaxy formation on gas clumping, we compare two sets of clusters simulated with the same initial conditions but with different prescription of gas physics. In the first set, we performed hydrodynamical cluster simulations without gas cooling and star formation. We refer this set of clusters as non-radiative (NR) clusters. In the second set, we turn on the physics of galaxy formation, such as metallicity-dependent radiative cooling, star formation, supernova feedback and a uniform UV background. We refer this set of clusters as cooling+star formation (CSF) clusters. We refer the reader to \citet{nagai_etal07a} for a detailed description of the gas physics implemented in these simulations. 

Our simulated sample includes 16 clusters at $z=0$ and their most massive progenitors at $z=0.6$. The properties of the simulated clusters at $z=0$ are given in Table~1 of \citet{lau_etal10}. In this work, we use the radius $r_{200}$ in which the enclosed mean overdensity is $200$ times the critical density at the redshift of the output. The cluster center is defined as the location of the most bounded dark matter particle. To study the dependence on the dynamical state of clusters, we distinguish unrelaxed and relaxed clusters based on the overall morphology of the mock X-ray images discussed in \cite{nagai_etal07b}. 

%%%%%%%%%%%%%%%%%%%%%%%%%%%%%
\begin{figure}[t]
\begin{center}
\vspace{-1cm}
\epsscale{1.2}
\plotone{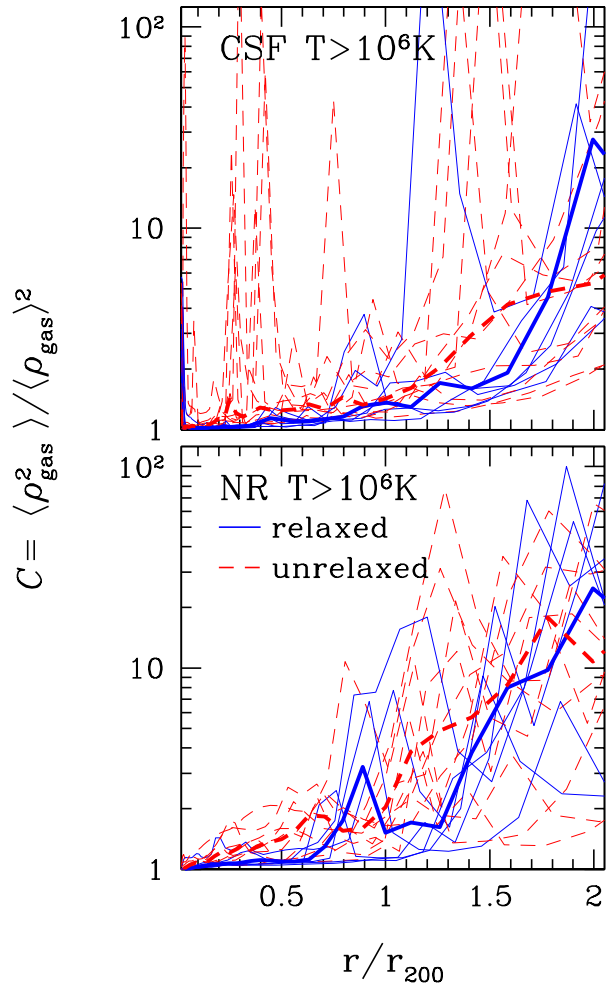}
\caption{The radial profiles of the clumping factor for the X-ray emitting gas with $T>10^6$~K.  We show results for the sample of 16 simulated clusters in the CSF (top) and NR (bottom) runs. Thin lines indicate individual profiles for relaxed (solid) and unrelaxed (dashed) clusters. Thick lines indicate the median profiles of each sub-sample.
}\label{fig:clump}
\end{center}
\end{figure}
%%%%%%%%%%%%%%%%%%%%%%%%%%%%%

%----------------------
\section{Clumping Factor}
\label{sec:clumping}
%----------------------

In the ICM, X-ray photons are emitted primarily by the scattering of electrons off ions via thermal bremsstrahlung. The X-ray surface brightness is given by 
\begin{equation}
S_X \propto \int n_e n_\H \Lambda(T,Z) \d V \propto \frac{1}{\mu_e\mu_\H m_\p^2}\int \rho_{\rm gas}^2  \Lambda(T,Z) \d V,
\label{eq:Sx}
\end{equation}
where $\Lambda(T,Z)$ is the cooling function which depends on temperature $T$ and metallicity $Z$, $\mu_e$ and $\mu_\H$ are the mean molecular weight of electrons and hydrogen, $m_{\rm p}$ is the proton mass, $\rho_\g$ is the gas density, and $V$ is the volume. Since $\Lambda(T,Z)$ depends weekly on $T$ and $Z \approx 0.3$, the observed X-ray surface brightness profile depends primarily on the average of gas density squared, 
\begin{equation}
S_X(r) \propto \langle \rho_{\rm gas}^2(r) \rangle = C(r) \langle \rho_{\rm gas}(r) \rangle^2,
\label{eq:Sx2}
\end{equation}
where we define the clumping factor as,
\begin{equation}
C\equiv\frac{\langle \rho_\g^2\rangle}{\,\langle \rho_\g\rangle^2} \geq 1.
\label{eq:clumping}
\end{equation}
Note that $C=1$, if the ICM is not clumpy (i.e., a single phase medium characterized by one temperature and gas density within each radial bin). 

In X-ray analyses of galaxy clusters, it is commonly assumed that the ICM is not clumpy. By setting $C=1$, the three dimensional gas density distribution is derived from the observed X-ray surface brightness profile by inverting Eq.~\ref{eq:Sx2}. However, if the ICM is clumpy, the gas density inferred from the X-ray surface brightness is overestimated by $\sqrt{C(r)}$. In what follows, we investigate the clumping factor of the X-ray emitting ICM using hydrodynamical cluster simulations.

%----------------------
\section{Results}
\label{sec:results}
%----------------------

%%%%%%%%%%%%%%%%%%%%%%%%%%%%%
\begin{figure}[t]
\begin{center}
\epsscale{1.2}
\plotone{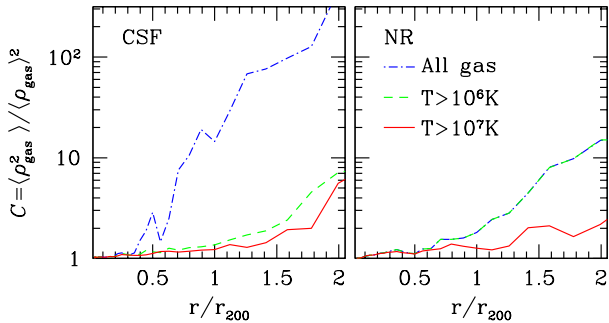}
\caption{ 
Median clumping factor profiles of gas with different minimum temperature in the CSF (left) and NR (right) runs. The dot-dashed line shows the median clumping factor profiles for all gas. Results for $T>10^6$~K and $T>10^7$~K are shown in dashed and solid lines, respectively.}\label{fig:clump_phase}
\end{center}
\end{figure}
%%%%%%%%%%%%%%%%%%%%%%%%%%%%%

%%%%%%%%%%%%%%%%%%%%%%%%%%%%%
\begin{figure}[t]
\begin{center}
\epsscale{1.2}
\plotone{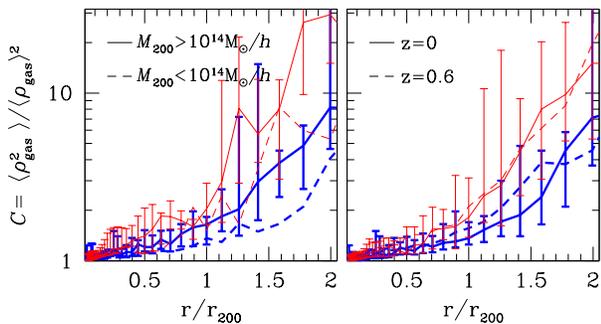}
\caption{ 
{\it Left}: the mass dependence of the median clumping factor profiles of gas with $T>10^6$~K. Results are shown for the present-day ($z=0$) clusters with $M_{200}>10^{14}\,h^{-1} M_{\odot}$ (solid lines) and $M_{200}<10^{14}\,h^{-1} M_{\odot}$ (dashed lines), respectively, in the CSF (thick) and NR (thin) runs. {\it Right}: the redshift dependence of the median clumping factor profiles in the CSF (thick) and NR (thin) runs at $z=0$ (solid lines) and $z=0.6$ (dashed lines). The errorbars on the solid lines indicate the interquartile range.
}\label{fig:clump_dep}
\end{center}
\end{figure}
%%%%%%%%%%%%%%%%%%%%%%%%%%%%%

\subsection{Clumping factor profile}
\label{subsec:clump}

In Figure~\ref{fig:clump} we plot the gas clumping factor as a function of radius for the sample of 16 simulated galaxy clusters.  Results for the CSF and NR runs are shown in the top and bottom panels respectively. Since we are primarily interested in the X-ray emitting ICM, we consider gas with temperature $T>10^6$~K here. We checked that the results based on $T>10^6$~K cut is essentially identical to the calculation that takes into account the energy-dependent effective area of the Suzaku XIS instrument. We thus apply the temperature cut of $T>10^6$~K when comparing to the X-ray data, unless stated otherwise. 

Figure~\ref{fig:clump} shows very large cluster-to-cluster variations, ranging between $C \sim 1-100$. As we illustrate below, these peaks correspond to high-density, low-temperature gas associated with infalling groups and galaxies. The amplitude of these peaks are generally more pronounced in the unrelaxed systems.  The scatter is larger in the CSF runs than in the NR runs, because gas condensation in infalling groups and satellites creates higher-density, lower-temperature gas that enhances the clumping factor. In the CSF runs, the clumping factor can exceed $C\gtrsim 100$ in regions with prominent clumps.

Since the scatter is highly non-Gaussian, we use the median (instead of the mean) for a measure of a {\it typical} clumping factor in our simulated clusters. In Figure~\ref{fig:clump}, we indicate the median clumping factor for the relaxed and unrelaxed systems using {\it thick-solid} and {\it thick-dashed} lines respectively. In both CSF and NR runs, the median clumping factor is approximately unity for $r<0.5r_{200}$, and increases gradually toward $r\sim r_{200}$.  At $r=r_{200}$, the clumping factor is $C\sim 1.3$ and $2$ for the CSF and NR runs respectively.  At $r>r_{200}$, the clumping factor increases rapidly with radius, reaching $C\sim 5$ and $10$ for the CSF and NR runs respectively. Note that the clumping factor of hot gas ($T\gtrsim 10^6$~K) is smaller in the CSF runs than in the NR runs, because some of the hot gas is cooled out of the X-ray band (see Figure~\ref{fig:clump_phase}). The clumping factor is generally larger in the unrelaxed systems, due to the enhanced level of substructures. Since the CSF runs suffer from the `overcooling' problem \citep[see e.g.,][]{kravtsov_etal05}, we argue that the real clumping factor should be bracketed by the results of our CSF and NR runs.

To investigate the dependence on gas temperature, we compute the clumping factor profiles for several different temperature thresholds. Figure~\ref{fig:clump_phase} shows the median clumping factor profiles calculated for gas with $T>10^6$~K and $T>10^7$~K, and for gas of all temperatures. For the X-ray emitting gas (e.g., $T> 10^6$~K), the clumping factor is considerably smaller in the CSF runs than in the NR runs.  In the CSF runs, the clumping factor is only $C\lesssim 2$ at $r=r_{200}$ and $C\sim 7$ at $r=2r_{200}$, which is less than the clumping in the NR runs. Raising the temperature threshold further to $T> 10^7$~K affects the clumping factor of the CSF run slightly, but affects the NR runs more dramatically, illustrating that significant fraction of clumping in the NR runs is due to the gas in the temperature range of $10^6<T<10^7$~K.  For gas of all temperatures, the clumping factor depends very sensitively on the cluster physics (e.g., cooling and star formation). In this case, the clumping factor is considerably larger in the CSF runs than in the NR runs. In the CSF runs, the clumping factor increases from $C\sim 2$ at $r=0.5 r_{200}$ to $C\sim 10$ at $r=r_{200}$ and $C\sim 100$ at  $r=2r_{200}$. In the NR runs, the clumping factor is comparable to that of the CSF runs at $r=0.5r_{200}$, but they are only $C\sim 2$ at $r=r_{200}$ and $C \sim 10$ at $r=2r_{200}$.  

Finally, to examine the dependence on the cluster mass and redshift, we compare the median clumping factor profiles for two mass bins and redshift bins for the gas with $T>10^6$~K in Figure~\ref{fig:clump_dep}.  In the left panel, we compare the sub-samples for $M_{200}>10^{14}\, h^{-1}$~M$_\odot$ and $M_{200}<10^{14}\, h^{-1}$~M$_\odot$. In the right panel, we compare the results for $z=0$ and $z=0.6$. In the CSF sample, lower mass systems show a smaller clumping factor, because compared to massive clusters, they have a larger fraction of lower temperature gas that are not detectable in the X-ray band. In the NR sample, this mass dependence is less pronounced. We find little evolution in the clumping factor profile for all runs. 

%%%%%%%%%%%%%%%%%%%%%%%%%%%%%
\begin{figure}[t]
\begin{center}
\epsscale{1.2}
\plotone{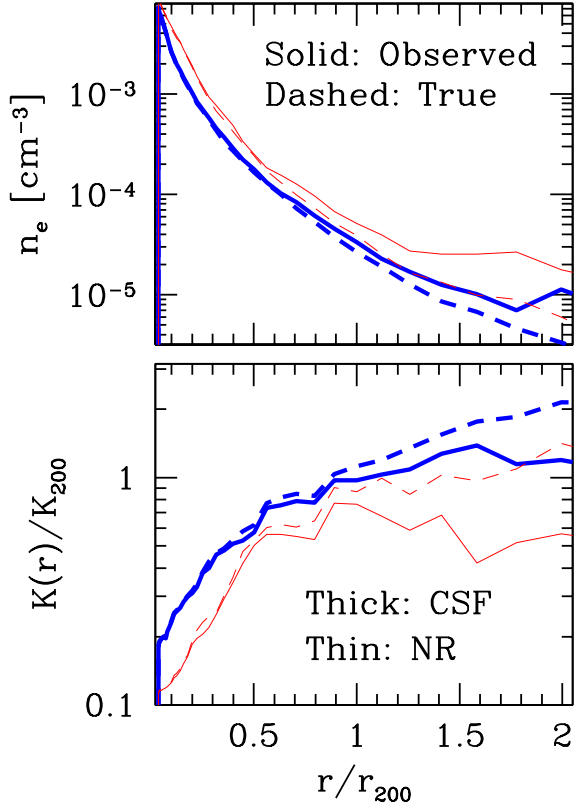}
\caption{ 
Effects of gas clumping on X-ray measurements of the ICM profiles. Gas clumping causes the overestimate of electron number density profile (top) and the flattening of the entropy profile (bottom) for gas with $T>10^6$~K. Solid lines indicate observed profiles, while dotted lines indicate true profiles. Thick and thin lines correspond to CSF and NR runs, respectively. 
}
\label{fig:pro}
\end{center}
\end{figure}
%%%%%%%%%%%%%%%%%%%%%%%%%%%%%

%%%%%%%%%%%%%%%%%%%%%%%%%%%%%
\begin{figure}[t]
\begin{center}
\epsscale{1.2}
\plotone{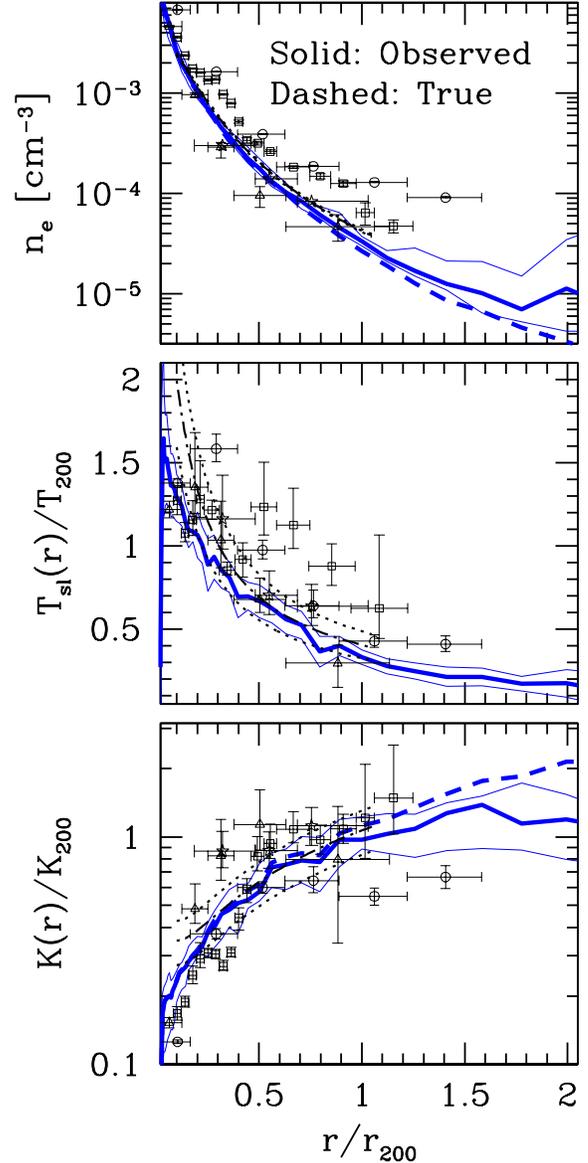}
\vspace{-0.8cm}
\caption{Comparisons of simulated clusters and Suzaku X-ray measurements. From top to bottom, we show the radial profiles for the electron number density, spectroscopic-like temperature, and entropy.  For the simulated clusters, we show the median profiles of the CSF runs. Dashed lines represent the true profiles. Solid lines represent the observed profiles, with thin-solid lines indicating the interquartile range. Black points are {\it Suzaku} observations of PKS0745-191 \citep{george_etal09} (circles), A1689 \citep{kawaharada_etal10} (triangles), A1413 \citep{hoshino_etal10} (stars), and Perseus East \citep{simionescu_etal11} (squares). The black dot-dashed lines are the best-fit profile of A1795 \citep{bautz_etal09} and the dotted black lines indicate $\pm 20\%$ uncertainty around the best-fit model. 
 }
\label{fig:pro_all}
\end{center}
\end{figure}
%%%%%%%%%%%%%%%%%%%%%%%%%%%%%

\subsection{Effects of Gas Clumping on the ICM profiles}
\label{subsec:pro}

In this section, we assess the effects of gas clumping on X-ray measurements of the ICM profiles. The {\it top} panel of Figure~\ref{fig:pro} shows that the observed and true median gas density profiles for the sample of 16 clusters at $z=0$. Results are shown for both CSF ({\it thick}) and NR ({\it thin}) runs. The solid lines indicate the {\it observed} gas density profiles, derived assuming $C(r)=1$. The dashed lines indicate the {\it true} profiles, defined as the volume-averaged gas density. At all radii, the observed gas density profile is higher by $\sqrt{C(r)}$ compared to the true profile.  In the CSF runs, the gas density is overestimated by 15\% at $r=r_{200}$ and 120\% at $r=2r_{200}$, respectively. Bias is larger for the NR runs, because of larger clumping factor (see Figures~\ref{fig:clump} and \ref{fig:clump_phase}): the observed gas density is higher by 40\% at $r=r_{200}$ and 300\% at $r=2r_{200}$, compared to the true value.  

Since the entropy of gas is defined as $K \equiv T/n_e^{2/3}$, the overestimate of gas density due to clumping causes an underestimate of the observed entropy profile by $C(r)^{1/3}$. This is shown in the {\it bottom} panel of Figure~\ref{fig:pro}, in which the entropy profiles are normalized to $K_{200}$ of the self-similar model \citep{kaiser86,voit05}. We use spectroscopic-like temperature $T_{\sl}$ \citep{mazzotta_etal04} to calculate entropy.\footnote{We only consider X-ray emitting gas with $T>10^6$~K when computing $T_{\sl}$.} The true entropy profile is consistent with the self-similar prediction with $K \propto r^{1.1}$ \citep{voit_etal05}. Since $C(r)$ is radially dependent, gas clumping causes flattening of the observed entropy profiles at $r\gtrsim r_{200}$. 

\subsection{Comparison with X-ray measurements}
\label{subsec:comp}

In Figure~\ref{fig:pro_all} we compare the observed gas density, temperature, and entropy profiles of our CSF runs to the {\it Suzaku} X-ray measurements of PKS0745-191 \citep{george_etal09}, Abell 1795 \citep{bautz_etal09}, Abell 1689 \citep{kawaharada_etal10}, Abell 1413 \citep{hoshino_etal10}, and Perseus East \citep{simionescu_etal11}. The temperature and entropy profiles are normalized to $T_{200}$ and $K_{200}$ from the self-similar model \citep[e.g.,][]{voit05}. 

The ``observed" ICM profiles of the simulated clusters are generally consistent with the {\it Suzaku} measurements, with the exception of PKS0745-191 which deviates significantly from our simulated clusters as well as other {\it Suzaku} measurements. Since the spectroscopic-like temperature profiles \citep{mazzotta_etal04} are in good agreement \citep[see also][]{burns_etal10}, the discrepancy in the entropy profile arises primarily from the difference in the gas density profile. 

While the flattening of the median profile of our simulated clusters is not as significant as the observed profile of PKS0745-191, we note that the simulated clusters show significant cluster-to-cluster variations, owing to the large variations in the clumping factor shown in Figure~\ref{fig:clump}). In fact, several simulated clusters exhibit the level of flattening similar to that of PKS0745-191 in their outskirts.  For example, as shown in Figure~\ref{fig:clump}, one of the relaxed clusters, classified as relaxed based on its morphology in the inner regions, exhibits a large clumping factor of $C\sim 100$ at $r=r_{200}$.  

However, we caution that these measurements are still quite uncertain. For example, it is puzzling that the gas density of PKS0745-191 is considerably larger than the other {\it Suzaku} measurements at all radii. Observations of more clusters and better understanding of systematic uncertainties are required before making robust conclusions.  Independent measurements with {\it Chandra}, {\it ROSAT},  {\it Swift}-XRT, and {\it XMM-Newton}, can shed light on systematic uncertainties in the current measurements \citep[e.g.,][]{ettori_molendi10, moretti_etal11}. 

%-----------------------------------------------%
\section{Conclusions and Discussions}
\label{sec:summary}
%-----------------------------------------------%

In this work, we show that gas clumping can cause significant biases in X-ray measurements of ICM profiles.  Using hydrodynamical cluster simulations, we show that gas clumping leads to an overestimate of the gas density and causes flattening of the entropy profile in the outskirts of galaxy clusters ($r \gtrsim r_{200}$).  

By comparing results of simulations with different input cluster physics, we find that the magnitude of gas clumping depends sensitively on the input cluster physics (e.g., gas cooling and star formation). For the X-ray emitting gas ($T\gtrsim 10^6$~K), the amount of gas clumping is smaller in clusters with radiative cooling and star formation (CSF) than those with simple non-radiative (NR) gas physics, because cooling removes some of the high-density, X-ray emitting gas out of the detectable temperature range. Our hydrodynamical simulations show that $C\sim 1.3$ and $2$ at $r=r_{200}$ and $C\sim 5$ and $10$ at $r= 2 r_{200}$ for the CSF and NR clusters, respectively. Since our CSF runs suffer from the overcooling problem, we argue that the real clumping factor should be bracketed by the results of our CSF and NR runs. We also checked that our results are essentially unchanged for the simulations for the WMAP-5 cosmology: $\Omega_M=1-\Omega_{\Lambda}=0.27$, $\Omega_B=0.0469$, $h=0.7$, $\sigma_8=0.82$.

The simulations also show that there is a large variation of the gas clumping factor among the cluster populations. The gas clumping is more pronounced in dynamically active systems, while it depends mildly on the cluster mass and redshift. Given the scatter, our results are consistent with the current observational constraints \citep[e.g.,][]{simionescu_etal11}. 

While current {\it Suzaku} measurements are still uncertain, the present work suggests that gas clumping is important for reducing the tension between the recent {\it Suzaku} observations and the theoretical predictions of the $\Lambda$CDM model.  We therefore caution that the effects of gas clumping must be taken into account when interpreting X-ray measurements of cluster outskirts. Results of this work will also be relevant for interpreting future X-ray missions, e.g., {\it eROSITA} and {\it WFXT}, which will produce large statistical samples of ICM measurements out to large cluster-centric radii. 

Further studies of gas clumping will be important for testing the predictions of hydrodynamical simulations and for using galaxy clusters as precision cosmological probes. Analyses of the nature of gas clumps (e.g., size distribution, filling factor, and angular dependence) using hydrodynamical simulations will be useful for designing observational strategies to constrain gas clumping. For example, gas clumping may be enhanced along the directions of filaments where the cluster accretes clumpy and diffuse materials, and it might help explain the anisotropic distribution of the ICM seen in recent simulations \citep[e.g.,][]{vazza_etal10} and in observations \citep[e.g.,][]{kawaharada_etal10}. Since the Sunyaev-Zel'dovich effect (SZE) signal is less sensitive to gas clumping, comparisons of X-ray and SZE measurements will shed light on the physics of cluster outskirts. 

\acknowledgements 
We thank Dominque Eckert, August Evrard, Matthew George, Alexey Vikhlinin, and the anonymous referee for comments on the manuscript. We also thank Matthew George for the data points for PKS0745-191 and Aurora Simionescu for Perseus. The cosmological simulations used in this study were performed on the IBM RS/6000 SP4 system (copper) at the National Center for Supercomputing Applications (NCSA). D.N. was supported in part by the NSF grant AST-1009811, by NASA ATP grant NNX11AE07G, and by the facilities and staff of the Yale University Faculty of Arts and Sciences High Performance Computing Center. 
\\

\end{document}